\documentclass[particles,article,accept,moreauthors]{Definitions/mdpi} 
\firstpage{1} 
\pubvolume{1}
\issuenum{1}
\articlenumber{0}
\pubyear{2024}
\copyrightyear{2024}
\datereceived{ }
\daterevised{ } 
\dateaccepted{ } 
\datepublished{ } 
\hreflink{https://doi.org/} 

\def \be  {\begin{equation}}
\def \ee  {\end{equation}}
\def \bea {\begin{eqnarray}}
\def \eea {\end{eqnarray}}
\def \nn  {\nonumber}
\newcommand{\de}{\partial}
\usepackage{amsmath}
\DeclareMathOperator{\Tr}{Tr}

\usepackage{mathrsfs}  

\usepackage[mathscr]{euscript}

\usepackage[makeroom]{cancel}

\usepackage[T1]{fontenc}
\usepackage{graphicx}

\Title{Derivation of Meson Masses in SU(3) and SU(4) Extended Linear-Sigma Model at Finite Temperature}

\TitleCitation{Derivation of Meson Masses in SU(3) and SU(4) Extended Linear-Sigma Model at Finite Temperature}

\Author{Abdel Nasser Tawfik $^{1,\ast}$ and Azar I. Ahmadov$^{2}$ and Alexandra Friesen $^{3}$, and Yuriy Kalinovsky $^{3}$, and Alexey Aparin $^{3}$, and Mahmoud Hanafy $^{4}$ } 

\AuthorNames{A. Tawfik, A. Ahmadov, A. Friesen, Yu. Kalinovsky, A. Aparin, and M. Hanafy}

\address{%
$^{1}$ \quad Future University in Egypt, Faculty of Engineering, Basic Science Department, 11835 New Cairo, Egypt \\ 
$^{2}$ \quad Institute for Physical Problems, Baku State University, Z. Khalilov st. 23, AZ1148, Baku, Azerbaijan \\ 
$^{3}$ \quad Joint Institute for Nuclear Research (JINR), 141980 Dubna, Russia Federation \\ 
$^{4}$ \quad Benha University, Faculty of Science, Physics Department, 13518 Benha, Egypt 
}

\corres{Correspondence: a.tawfik@fue.edu.eg}

\abstract{
The present study focuses on the mesonic potential contributions to the Lagrangian of the extended linear-sigma model (eLSM) for scalar and pseudoscalar meson fields across various quark flavors. The present study focuses on the low-energy phenomenology associated with quantum chromodynamics (QCD), where mesons and their interactions serve as the pertinent degrees of freedom, rather than the fundamental constituents of quarks and gluons. Given that SU(4) configurations are completely based on SU(3) configurations, the possible relationships between meson states in SU(3) and those in SU(4) are explored at finite temperature. Meson states, which are defined by distinct chiral properties, are grouped according to their orbital angular momentum $J$, parity $P$, and charge conjugation $C$. Consequently, this organization yields scalar mesons with quantum numbers $J^{PC}=0^{++}$, pseudoscalar mesons with $J^{PC}=0^{-+}$, vector mesons with $J^{PC}=1^{--}$, and axialvector mesons with $J^{PC}=1^{++}$. We accomplished the derivation of analytical expressions for a total of seventeen noncharmed meson states and twenty-nine charmed meson states so that an analytical comparison of the noncharmed and charmed meson states at different temperatures becomes feasible and the contributions SU(3) and SU(4) configurations can be estimated, analytically.
}

\begin{document}

\setcounter{section}{-1} 

\section{Introduction}
\label{sec:intr}

Perturbative approaches to Quantum Chromodynamics (QCD) yield effective solutions, particularly at very high energy scales \cite{Ferreira:2023fva}. In contrast, nonperturbative solutions require numerical methods for approximation \cite{Gorenstein:1988fe}. The substantial computational costs associated with these methods highlight the necessity for accurate solutions at nonzero density. Nevertheless, numerical lattice QCD simulations face limitations at finite density; the "sign problem" \cite{Danzer:2009dk}. This situation has led to the adoption of effective models \cite{Radzhabov:2010dd,Contrera:2012wj}, such as the hadron resonance gas model \cite{Karsch:2003vd,Karsch:2003zq,Tawfik:2004vv,Tawfik:2004sw}. As for QCD-like effective models, Nambu--Jona--Lasinio (NJL) model appears as a widely used approach \cite{Coppola:2018vkw,Carlomagno:2022inu,Carlomagno:2022arc}, which was introduced to investigate the source of nucleon mass as a self-energy within a framework characterized by four-fermion interactions, drawing parallels to the emergence of an energy gap in superconductivity theory \cite{PhysRev.122.345}. This model serves as a phenomenological representation of quarks, exhibiting chiral symmetry breaking at low densities and temperatures, while restoring chiral symmetry at elevated densities and temperatures. In the phase where chiral symmetry is broken, quarks acquire a dynamical mass through their interactions with the vacuum. Furthermore, QCD-like models, including the extended linear-sigma model (eLSM) introduced in the 1960s, are significant due to their shared global symmetries with QCD and their lower computational requirements \cite{Gausterer:1988fv,Friesen:2018ojv,Kalinovsky:2015kzf,Friesen:2011wt}. In this context, it is important to note that the Lagrangian associated with the gauge theory governing the color interactions of quarks and gluons exhibits invariance under local color transformations. This property ensures that the physical implications remain unchanged when the colors of quarks and gluons undergo transformation. Conversely, the interactions themselves do not depend on flavor. In the effective QCD-model, eLSM, the global chiral symmetry is explicitly broken due to the presence of non-zero quark masses and quantum effects, as discussed in the literature \cite{Giacosa2012}. Additionally, this symmetry is spontaneously broken by the non-zero expectation value of the quark condensate within the QCD vacuum.

The eLSM is utilized in exploring various QCD symmetries. This includes i) isospin symmetry, which is characterized as a global transformation linked to SU(2) rotations in the flavor space of up and down quarks, where the Lagrangian remains invariant for equal or negligible masses \cite{Tawfik:2019tkp,Tawfik:2023egf,Tawfik:2023all}. Moreover, ii) global chiral symmetry is maintained in the chiral limit of massless QCD, where left- and right-handed components exhibit symmetry. Nonetheless, this symmetry is violated due to the properties of the QCD vacuum, including the Higgs mechanism, which leads to pions being recognized as the lightest Goldstone bosons resulting from the broken symmetry \cite{Tawfik:2015uda,Tawfik:2015tga}. The eLSM investigation also addresses iv) discrete symmetries such as charge conjugation (C), parity (P), and time reversal (T) \cite{Eshraim2015}, along with v) classical dilatation (scale) symmetry \cite{Eshraim2020}.

Before the establishment of QCD, which serves as the theoretical foundation for strong interactions, Gell--Mann and Levy proposed the sigma model to introduce a field associated with a point particle. This field is restricted to a defined manifold and characterizes the interactions of pions \cite{Gell-Mann:1960mvl}. The field corresponds to a spinless sigma, referred to as the scalar meson. A variety of studies have utilized the LSM, including the $\mathcal{O}(4)$ LSM \cite{Gell-Mann:1960mvl}. The implementation of LSM, which incorporates quark degrees of freedom, is enabled by an extension that realizes the dynamic nature of pseudoscalar and scalar mesons as a linear representation of chiral symmetry, which experiences weak breaking due to the current quark masses. This study is dedicated to exploring the low-energy phenomenology of quantum chromodynamics (QCD) \cite{Burkert:2022hjz}, emphasizing mesons and their interactions as the relevant degrees of freedom, in contrast to the fundamental elements of quarks and gluons. Thus, the model can be regarded as an effective representation of nonperturbative QCD \cite{Lenaghan:1999si,Petropoulos:1998gt,PhysRevD.9.1825,PhysRevD.12.2781,PhysRevD.21.278} that helps in understanding the emergence of hadron masses and structures \cite{Ding:2022ows}. 

Chiral symmetry is commonly viewed as a foundational approximation for understanding the structure of hadrons. Approximately forty years ago, calculations of the spin-zero mass spectrum and leptonic decay constants were performed using the one-loop approximation of the SU(4) linear-sigma model \cite{Geddes:1979nd}. The extended linear sigma model (eLSM) investigated the phenomenological aspects of charmed mesons \cite{Eshraim2015}. More recently, the quark-hadron phase diagram has been examined at constrained temperatures and densities through the mean-field approximation of the SU(4) Polyakov linear-sigma model (PLSM) \cite{Diab:2016iig,AbdelAalDiab:2018hrx}. It is proposed that the $N_f=4$ Lagrangian bears similarities to the $N_f=3$ Lagrangian. In a thorough investigation, one of the contributing authors, AT, utilized SU(3) PLSM to analyze the thermodynamic behavior, phase structure, and meson masses of QCD across a range of finite temperatures, densities, and magnetic fields. This extensive research \cite{Tawfik:2014uka,Tawfik:2014gga,Tawfik:2016lih,Tawfik:2016gye,Tawfik:2016ihn,Tawfik:2016edq,Tawfik:2017cdx,Tawfik:2019rdd,Tawfik:2019tkp,Tawfik:2019kaz,Tawfik:2021eeb} provided critical insights into various fundamental features of QCD.

The present study utilizes eLSM) that features two configurations. The first configuration includes three flavors of quarks, while the second configuration incorporates an additional charm quark flavor. Each configuration is combined with scalar and pseudoscalar meson fields \cite{Diab:2016iig,AbdelAalDiab:2018hrx,Klempt:2007cp,AMSLER200461}. To simplify our discussion, we will analyze the contributions of each configuration to the meson potential. The quark sets enable an analytical evaluation of various meson states, expressed as $\langle \bar{q} q\rangle = \langle \bar{q}_{\ell} q_{r} - \bar{q}_{\ell} q_{r}\rangle \neq 0$ \cite{VAFA1984173}. Meson states, which exhibit distinct chiral structures, are organized according to certain quantum numbers, including orbital angular momentum $J$, parity $P$, and charge conjugation $C$. This organization results in the classification of scalar mesons as $J^{PC}=0^{++}$, pseudoscalar mesons as $J^{PC}=0^{-+}$, vector mesons as $J^{PC}=1^{--}$, and axialvector mesons as $J^{PC}=1^{++}$ \cite{Tawfik:2019rdd,Tawfik:2014gga}. We have compiled analytical analyses on seventeen noncharmed meson states and twenty-nine charmed meson states at finite temperature. The focus of this study is to undertake an analytical comparison of the noncharmed and charmed meson states at different temperatures. 
The analysis of density dependence could be addressed in a separate manuscript. The fundamental argument suggests that density should be depicted in a manner comparable to that of temperature.

The script is structured in the following manner: Section \ref{sec:frml} introduces the underlying formalism. The details of the SU(3) configuration and the seventeen noncharmed meson states are provided in section \ref{sec:su3}. Section \ref{sec:su4} presents the SU(4) configuration along with the twenty-nine charmed meson states. Lastly, section \ref{sec:cncl} is devoted to the conclusions and an outlook on future research.

\section{Mesonic Potential of extended Linear-Sigma Model}
\label{sec:frml}

The sigma model integrates chiral symmetry through a linear representation \cite{Gell-Mann:1960mvl}. Unlike the nonlinear representation, which addresses only the Goldstone bosons and omits vector mesons, the linear representation permits a comprehensive study of both Goldstone bosons and scalar mesons. This expansion into the vector sector allows for the inclusion of vector and axialvector mesons. The eLSM acknowledges chiral symmetry in conjunction with other QCD symmetries \cite{Rosenzweig:1981cu}. In section \ref{sec:su3}, we will introduce the SU(3) mesonic potential. Comprehensive information regarding the complete Lagrangian is available in Ref. \cite{Ahmadov:2023mmy}. Moreover, we will introduce an analytical description for seventeen meson states, expanding our discussion to encompass finite temperature considerations. This analysis aims to assess the influence of finite temperature on the analytical expressions related to meson states.

\subsection{SU(3) Configuration}
\label{sec:su3}
Let us start by introducing the SU(3) mesonic potential
\bea
U(\sigma_x,\sigma_y) &=& \frac{m^2}{2}\left(\sigma_x^2+\sigma_y^2\right) - \frac{\mathscr{C}}{2\sqrt{2}} \sigma_x^2 \sigma_y +\frac{\ell_1}{2} \sigma_x^2\sigma_y^2 + \frac{1}{8} \left(2\ell_1+\ell_2\right)\sigma_x^4 \nn \\
&+& \frac{1}{4} \left(\ell_1+\ell_2\right) \sigma_y^4 - h_x \sigma_x - h_y \sigma_y. \label{eq:PotSU3}
\eea
The minimal global minimization of the grand potential allows for the deduction of the light quark condensates, denoted as $\sigma_x$ and $\sigma_y$, which represent the light and strange quark condensates, respectively. 

All condensates and parameters in the effective mesonic potential equation Eq. (\ref{eq:PotSU3}) shall be regarded as functions of temperature. The temperature dependence of $T$, $m$, $h_x$ and $h_y$ are explicitly given in \eqref{eq:m}, \eqref{eq:T}, \eqref{eq:hx}, and \eqref{eq:hy}, respectively. To examine the temperature dependence of the various quantities associated with eLSM, we begin with:
\bea
m^2(T) &=& m^2\left(1-\frac{T^2}{T_0^2}\right), \quad \mathtt{where} \;\; T_0 \simeq \Lambda_{\mathtt{QCD}}, \label{eq:m} \\
T_{N_c}(T) &=& \frac{T_0}{\sqrt{1+\frac{T^2}{2f_{\pi}^2} \frac{3}{N_c}}}, \quad \mathtt{where} \;\;  \lim_{N_c\rightarrow\infty} T_{N_c} = T_0, \label{eq:T}
\eea
where $T_0$ is the deconfinement critical temperature \cite{Bowman:2008kc,Tawfik:2016gye,Tawfik:2015uda} and $N_c$ refers to the color degrees of freedom found in non-Abelian SU(N$_c$) gauge theory \cite{Datta:2009jn}. Additionally, $T_{N_c}(T)$ is the critical temperature that characterizes gauge QCD, which includes color degrees of freedom, and $\Lambda_{\mathtt{QCD}}$ denotes the energy scale associated with QCD \cite{ParticleDataGroup:2022pth}.

The anomaly breaking term $\mathscr{C}$ associated with U(1)$_{\mathtt{A}}$ is constrained by the function $\ell_2(T)$ alongside the mass difference observed between pions and kaons:
\bea
\mathscr{C}(T) &=& \frac{m_K^2(T) - m_{\pi}^2(T)}{f_K-f_{\pi}} - \ell_2\left(2 f_K-f_{\pi}\right).
\eea 

At the conditions where $\frac{\partial U}{\partial \sigma_x} = 0$ and $\frac{\partial U}{\partial \sigma_y} = 0$, the quantities $h_x(T)$ and $h_y(T)$ can be analytically determined. 
\bea
h_x(T) &=& m^2(T) \sigma_x(T) - \frac{\mathscr{C}}{\sqrt{2}} \sigma_x(T) \sigma_y(T) - \ell_1(T) \sigma_x(T) \sigma_y^2(T) + \frac{1}{2}\left[2\ell_1(T)+\ell_2(T)\right] \sigma_x^3(T), \hspace*{7mm} \label{eq:hx}\\
h_y(T) &=& m^2(T) \sigma_x(T) - \frac{\mathscr{C}}{2\sqrt{2}} \sigma_x^2(T) - \ell_1(T) \sigma_x^2(T) \sigma_y(T) + \left[\ell_1(T)+\ell_2(T)\right] \sigma_y^3(T). \label{eq:hy}
\eea

The dependence of $\ell_1$ and $\ell_2$ on temperature is represented as follows:
\bea
\ell_1(T) &=& \frac{m_{\sigma}^2(T) - m_{\pi}^2(T) - m_{a_0}^2(T) + m^2_{\eta}(T)}{3f_{\pi}^2}, \\
\ell_2(T) &=& \frac{3\left[2f_{K}-f_{\pi}\right]m_K^2(T) - 3\left[2f_{K}-f_{\pi}\right]m_{\pi}^2(T) - 2\left[f_{K}-f_{\pi}\right] \left[m_{\eta^{\prime}}^2(T) + m_{\eta}^2(T)\right]}{\left[f_{K}-f_{\pi}\right]\left[3f_{\pi}^2+8f_K\left(f_K-f_{\pi}\right)\right]}, \hspace*{7mm}
\eea
where $f_{\pi}$ is the pion decay constant \cite{Gasser:2010wz} and $f_{K}$ is the kaon decay constant \cite{Sanz-Cillero:2004hed}. Moreover, within this configuration, the masses corresponding to the meson states, namely $m_{\pi}(T)$, $m_K(T)$, $m_{\sigma}(T)$, $m_{a_0}(T)$, $m_{\eta}(T)$, and $m_{\eta^{\prime}}(T)$, are to be calculated.

Now, the formulation of analytical expressions regarding the masses of seventeen noncharmed mesons, emphasizing their relationship with finite temperatures, shall be outlined. 
The grand potential can be derived in the mean field approximation. Given the assumption of thermal equilibrium, the grand partition function is expressed via a path integral that includes the quark, antiquark, and meson fields
\begin{eqnarray}
\mathcal{Z}&=& \mathrm{Tr \,exp}[-\hat{\mathcal{H}}/T] = \int\prod_a \mathcal{D} \sigma_a \mathcal{D} \pi_a \int
\mathcal{D}\psi \mathcal{D} \bar{\psi} \mathrm{exp} \left[\int_x \mathcal{L}\right], 
%
\end{eqnarray}
where $\int_x\equiv i \int^{1/T}_0 dt \int_V d^3x$, $t$ is the time, at which the system with volume $V$ evolves. The present analysis assumes vanishing density, which corresponds to vanishing chemical potential. The partition function can be derived using the mean field approximation \cite{Schaefer:2008hk,Schaefer:2006ds,Scavenius:2000qd}. In this context, the meson fields are substituted with their expectation values, specifically $\bar{\sigma_x}$ and $\bar{\sigma_y}$, within the action \cite{Cheng:2007jq,Kapusta:2006pm}. By employing standard techniques \cite{Kapusta:2006pm}, we can perform the integration over the fermionic contributions, leading to the derivation of the effective potential for the mesons
\begin{eqnarray}
\Omega(T)=\frac{-T \mathrm{ln}
\mathcal{Z}}{V}=U(\sigma_x, \sigma_y)+\mathbf{\mathcal{U}}(\Phi, \Phi^*, T)+\Omega_{\bar{\psi}
\psi}, \label{potential}
\end{eqnarray}
where the fields $\Phi$ or $\Phi^*$ are defined, Eq. \eqref{eq:PHI},  as a complex matrix of dimensions $N_f \times N_f$, which includes the scalar $\sigma_a$ with quantum numbers $J^{PC}=0^{++}$, the pseudoscalar $\pi_a$ with $J^{PC}=0^{-+}$, vector mesons that possess $J^{PC}=0^{--}$, and axialvector mesons that share the quantum numbers $J^{PC}=0^{++}$. The mesonic potential is explicitly elaborated herein. Additional information on the other potentials $\mathbf{\mathcal{U}}(\Phi, \Phi^*, T)$ and $\Omega_{\bar{\psi}\psi}$ can be found in previously published works \cite{Tawfik:2014uka,Tawfik:2014gga,Tawfik:2016lih,Tawfik:2016gye,Tawfik:2016ihn,Tawfik:2016edq,Tawfik:2017cdx,Tawfik:2019rdd,Tawfik:2019tkp,Tawfik:2019kaz,Tawfik:2021eeb}.

By second derivation of the grand potential evaluated at its minimum with respect to the corresponding fields, the masses of various states can be obtained. In the present calculations, the minima are estimated by vanishing expectation values of all scalar, pseudoscalar, vector and axialvector fields. 
\bea
m^2_{i, ab} &=& \left.\frac{\de^2 \Omega(T)}{\de \beta_{i,a} \de \beta_{i,b}}\right|_{\mathtt{min}}, \label{eq:GrndPtnSU3}
\eea
where $\beta_{i,a}$ and $\beta_{i,b}$ are the corresponding mass fields of $i$-th hadron state. $a, b\in[0,1,\cdots,8]$.

\begin{itemize}
\item Scalar noncharmed mesons:
\bea
m^2_{\sigma_N}(T) &=& m^2_0\left(1-\frac{T^2}{T_0^2}\right) + \frac{3}{2} \ell_2(T) \sigma^2_x(T), \\
m^2_{K_0^{\ast}}(T) &=& Z^2_{K_0^{\ast}}(T) \left[m^2_0\left(1-\frac{T^2}{T_0^2}\right) + \frac{\ell_2(T)}{2} \sigma^2_x(T) \right. \nn \\
&&\left. + \frac{\ell_2(T)}{\sqrt{2}} \sigma_x(T) \sigma_y(T) + \ell_2(T) \sigma^2_y(T)\right], \hspace*{7mm} \\
m^2_{a_0}(T) &=& m^2_0\left(1-\frac{T^2}{T_0^2}\right) + \frac{3}{2} \ell_2(T) \sigma^2_x(T), \\
m^2_{\sigma_S}(T) &=& m^2_0\left(1-\frac{T^2}{T_0^2}\right) + 3 \ell_2(T) \sigma^2_y(T).
\eea
\item Pseudoscalar noncharmed mesons:
\bea
m^2_{\pi}(T) &=& Z^2_{\pi}(T) \left[m^2_0\left(1-\frac{T^2}{T_0^2}\right) + \frac{\ell_2(T)}{2} \sigma^2_x(T)\right], \\
m^2_{K}(T) &=& Z^2_{K}(T) \left[m^2_0\left(1-\frac{T^2}{T_0^2}\right) + \frac{\ell_2(T)}{2} \sigma^2_x(T) \right. \nn \\
&& \left. - \frac{\ell_2(T)}{\sqrt{2}} \sigma_x(T) \sigma_y(T) + \ell_2(T) \sigma^2_y(T)\right], \hspace*{7mm} \\
m^2_{\eta_N}(T) &=& Z^2_{\pi}(T) \left[m^2_0\left(1-\frac{T^2}{T_0^2}\right) + \frac{\ell_2(T)}{2} \sigma^2_x(T) + \mathscr{D} \sigma^2_x(T) \sigma^2_y(T) \right], \\
m^2_{\eta_S}(T) &=& Z^2_{\eta_S}(T) \left[m^2_0\left(1-\frac{T^2}{T_0^2}\right) + \ell_2(T) \sigma^2_y(T) + \frac{\mathscr{D}}{4} \sigma^4_x(T) \right], \\
m^2_{\eta_{NS}}(T) &=& Z_{\pi}(T) Z_{\pi_3}(T) \frac{\mathscr{D}}{2} \sigma^3_x(T) \sigma_y(T).
\eea
\item Axialvector noncharmed mesons:
\bea
m^2_{a_1}(T) &=& m^2_1(T) - m^2_0\frac{T^2}{T_0^2} + \frac{1}{2} \left(2 g_1^2(T) + h_2(T) - h_3(T)\right) \sigma_x^2(T), \\
m^2_{K_1}(T) &=& m^2_1(T) - m^2_0\frac{T^2}{T_0^2} + \frac{1}{4} \left(g_1^2(T) + h_2(T)\right) \sigma_x^2(T) - \frac{1}{\sqrt{2}}  \sigma_x(T)  \sigma_y \left(h_3(T)-g_1^2(T)\right) \nn \\
&+& \frac{1}{2}\left(h_2(T)+g_1^2(T)\right) \sigma^2_y(T) + \delta_s(T), \\
m^2_{f_{1S}}(T) &=& m^2_1(T) - m^2_0\frac{T^2}{T_0^2} + \left(2 g_1^2(T) + h_2(T) - h_3(T)\right) \sigma_y^2(T) + 2 \delta_s(T),\\
m^2_{f_{1N}}(T) &=& m^2_{a_1}(T).
\eea
\item Vector noncharmed mesons:
\bea
m^2_{\rho}(T) &=& m^2_1(T) - m^2_0\frac{T^2}{T_0^2} + \frac{1}{2}\left(h_2(T) + h_3(T)\right) \sigma_x^2(T),\\
m^2_{K^{\ast}}(T) &=& m^2_1(T) - m^2_0\frac{T^2}{T_0^2} + \frac{1}{4}\left(g_1^2(T) + h_2(T)\right) \sigma_x^2(T) + \frac{1}{\sqrt{2}}  \sigma_x(T) \sigma_y \left(h_3(T)-g_1^2(T)\right) \nn \\
&+& \frac{1}{2}\left(h_2(T)+g_1^2(T)\right) \sigma^2_y(T) + \delta_s(T),\\
m^2_{\omega_{S}}(T) &=& m^2_1(T) - m^2_0\frac{T^2}{T_0^2} + \left(h_2(T) + h_3(T)\right) \sigma_y^2(T) + 2 \delta_s(T),  \\
m^2_{\omega_{N}}(T) &=& m^2_{\rho}(T).
\eea
\end{itemize}

The different quantities present in the mass expressions are outlined as follows: \bea
g_1^2(T) &=& \frac{m_{a_1}^2(T)}{f_{\pi}^2 Z_{\pi}^2} \left(1-\frac{1}{Z_{\pi}^2}\right), \\
\mathscr{D}(T) &=& \frac{1}{2} \left(m^2_{\eta}(T) - m^2_{\pi}(T)\right), \\
\delta_s(T) &=& \frac{1}{2}\left\{m^2_{\omega_S}(T) - m^2_1(T) + m^2_0 \frac{T^2}{T^2_0} - \left[\frac{h_1^2(T)}{2} + h_2(T) + h_3(T)\right] \sigma^2_y(T) \right. \nn \\
&-&\left.  \frac{h_1(T)}{2} \sigma^2_x(T)\right\}, \\
m^2_1(T) &=& m^2_{\omega_S}(T) - \left[\frac{h_1(T)}{2} + h_2(T) - h_3(T)\right] \sigma^2_s(T) - \frac{h_1(T)}{2} \sigma_s(T) - 2 \delta_s(T), \\
Z_{\pi}(T) &=& Z_{\eta_N}(T) = \frac{m_{a_1}(T)}{\sqrt{m^2_{a_1}(T)-g^2_1(T) \sigma^2_x(T)}}, \\
Z_K(T) &=& \frac{2 m_{K_1}(T)}{4 m^2_{K_1}(T) - g^2_1(T)\left[\sigma_x(T) + \sqrt{2} \sigma_y(T)\right]^2},\\
Z_{\eta_S}(T) &=& \frac{m_{f_{1S}}(T)}{m^2_{f_{1S}}(T) - 2 g^2_1(T) \sigma^2_y(T)}, \\
Z_{K^{\ast}_0}(T) &=& \frac{2 m_{K^{\ast}_0}(T)}{4 m^2_{K^{\ast}_0}(T) - g^2_1(T) \left[\sigma_x(T) - \sqrt{2} \sigma_y(T)\right]^2}.
\eea

Section \ref{sec:su4} introduces further analytical representations of meson states. This discussion specifically emphasizes charmed meson states, which exhibit variations in response to changes in temperature.

\subsection{SU(4) Configuration}
\label{sec:su4}

It is assumed that the Lagrangian for $N_f=4$, which maintains global chiral invariance \cite{Diab:2016iig}, is similar to that of $N_f=3$ \cite{Parganlija:2012fy}. In the context of $N_f=4$, the mass term $- 2 \Tr[\epsilon \Phi^{\dagger} \Phi]$ must be integrated into the eLSM Lagrangian \cite{Giacosa2012}. The corresponding mesonic grand potential is given as 
\bea
U_m^{SU(4)}(\Phi) &=& \frac{m^2}{2} \left(\sigma_x^2+\sigma_y^2+\sigma_{15}^2\right) + \ell_1\left[4\left(\sigma_x+\frac{\sigma_{15}}{\sqrt{6}}\right)^4 + \left(\sqrt{2}\sigma_y+\frac{\sigma_{15}}{\sqrt{6}}\right)^4 \right. \nn \\
&&\left. + \left(\sqrt{\frac{2}{3}} \sigma_0 - \sqrt{\frac{3}{2}}\sigma_{15}\right)^4 + 4\left(\sigma_x+\frac{\sigma_{15}}{\sqrt{6}}\right)^2 \left(\sqrt{2}\sigma_y+\frac{\sigma_{15}}{\sqrt{6}}\right)^2 \right. \nn \\
&&\left. 
+ 4\left(\sigma_x+\frac{\sigma_{15}}{\sqrt{6}}\right)^2   \left(\sqrt{\frac{3}{2}}\sigma_0-\sqrt{\frac{3}{2}} \sigma_{15}\right)^2  + 2 \left(\sqrt{2}\sigma_y+\frac{\sigma_{15}}{\sqrt{6}}\right)^2 \left(\sqrt{\frac{2}{3}}\sigma_0-\sqrt{\frac{3}{2}} \sigma_{15}\right)^2
\right] \nn \\
&+& \ell_2\left[2\left(\sigma_x+\frac{\sigma_{15}}{\sqrt{6}}\right)^4 + \left(\sqrt{2}\sigma_y+\frac{\sigma_{15}}{\sqrt{6}}\right)^4 + \left(\sqrt{\frac{2}{3}} \sigma_0-\sqrt{\frac{3}{2}}\sigma_{15}\right)^4
\right] \nn \\
&-& \frac{c}{8} \left[\frac{2}{3} \sigma_x^2 \sigma_y \sigma_0 + \frac{\sigma_y \sigma_{15}^2 \sigma_0}{3\sqrt{3}} + \frac{2\sqrt{2}}{3}\sigma_x \sigma_y \sigma_{15} \sigma_0 + \frac{1}{3}\sigma_x^2 \sigma_{15} \sigma_0 + \frac{\sigma_{15}^3 \sigma_0}{8} \right. \nn \\
&& \left. + \frac{\sqrt{2}\sigma_x\sigma_{15}\sigma_0^2}{3\sqrt{3}} - \sqrt{3} \sigma_x^2 \sigma_y \sigma_{15} - \frac{\sigma_{15}^3 \sigma_y}{2\sqrt{3}} - \sqrt{2}\sigma_x\sigma_y\sigma_{15}^2 - \frac{\sigma_x^2\sigma_{15}^2}{2} - \frac{\sigma_{15}^4}{12} - \frac{\sigma_x \sigma_{15}^3}{\sqrt{6}}
\right]. \label{eq:PntlMsnSU4}
\eea The field $\Phi$ is given as
\bea
\Phi &=& \sum_{a=0}^{N_f^2-1} T_a \left(\sigma_a + i \pi_a\right), \label{eq:PHI}
\eea
where the scalar mesons are given as {\footnotesize
\bea
T_a \sigma_a &=& \frac{1}{\sqrt{2}} \nn \\
&& \left(
\begin{tabular}{c c c c}
$\frac{\sigma_0}{2}+\frac{\sigma_3}{\sqrt{2}}+\frac{\sigma_8}{\sqrt{6}}+\frac{\sigma_{15}}{2\sqrt{3}}$ & $\frac{\sigma_1-i\sigma_2}{\sqrt{2}}$ & $\frac{\sigma_4-i\sigma_{5}}{\sqrt{2}}$ & $\frac{\sigma_9-i\sigma_{10}}{\sqrt{\sqrt{2}}}$ \\
$\frac{\sigma_1+i\sigma_2}{\sqrt{2}}$ & $\frac{\sigma_0}{\sqrt{2}}-\frac{\sigma_3}{\sqrt{2}}+\frac{\sigma_8}{\sqrt{6}}+\frac{\sigma_{15}}{2\sqrt{3}}$ & $\frac{\sigma_6-i\sigma_7}{\sqrt{2}}$ & $\frac{\sigma_{11}-i\sigma_{12}}{\sqrt{2}}$ \\
$\frac{\sigma_4+i\sigma_5}{\sqrt{2}}$ & $\frac{\sigma_6+i\sigma_7}{\sqrt{2}}$ & $\frac{\sigma_0}{2}-\sqrt{\frac{2}{3}}\sigma_8 + \frac{\sigma_{15}}{2\sqrt{3}}$ & $\frac{\sigma_{13}-i\sigma_{14}}{\sqrt{2}}$ \\
$\frac{\sigma_9+i\sigma_{10}}{\sqrt{2}}$ & $\frac{\sigma_{11}+i\sigma_{12}}{\sqrt{2}}$ & $\frac{\sigma_{13}+i\sigma_{14}}{\sqrt{2}}$ & $\frac{\sigma_0}{2}-\frac{\sqrt{3}}{2}\sigma_{15}$
\end{tabular}
\right). \hspace*{10mm} \label{eq:Tapia}
\eea }
In a similar manner, the pseudo-scalar mesons are given as {\footnotesize
\bea
T_a \pi_a &=& \frac{1}{\sqrt{2}}  \nn \\
&& \left(
\begin{tabular}{c c c c}
$\frac{\pi_0}{2}+\frac{\pi_3}{\sqrt{2}}+\frac{\pi_8}{\sqrt{6}}+\frac{\pi_{15}}{2\sqrt{3}}$ & $\frac{\pi_1-i\pi_2}{\sqrt{2}}$ & $\frac{\pi_4-i\pi_{5}}{\sqrt{2}}$ & $\frac{\pi_9-i\pi_{10}}{\sqrt{\sqrt{2}}}$ \\
$\frac{\pi_1+i\pi_2}{\sqrt{2}}$ & $\frac{\pi_0}{\sqrt{2}}-\frac{\pi_3}{\sqrt{2}}+\frac{\pi_8}{\sqrt{6}}+\frac{\pi_{15}}{2\sqrt{3}}$ & $\frac{\pi_6-i\pi_7}{\sqrt{2}}$ & $\frac{\pi_{11}-i\pi_{12}}{\sqrt{2}}$ \\
$\frac{\pi_4+i\pi_5}{\sqrt{2}}$ & $\frac{\pi_6+i\pi_7}{\sqrt{2}}$ & $\frac{\pi_0}{2}-\sqrt{\frac{2}{3}}\pi_8 + \frac{\pi_{15}}{2\sqrt{3}}$ & $\frac{\pi_{13}-i\pi_{14}}{\sqrt{2}}$ \\
$\frac{\pi_9+i\pi_{10}}{\sqrt{2}}$ & $\frac{\pi_{11}+i\pi_{12}}{\sqrt{2}}$ & $\frac{\pi_{13}+i\pi_{14}}{\sqrt{2}}$ & $\frac{\pi_0}{2}-\frac{\sqrt{3}}{2}\pi_{15}$
\end{tabular}
\right). \hspace*{10mm} \label{eq:Tasigmaa}
\eea
}

It is apparent that the global minima, defined by the absence of partial derivatives related to $\sigma_x$, $\sigma_y$, and $\sigma_c$, give rise to
\bea
h_x &=& m^2 \sigma_x - \frac{\mathscr{C}}{2}  \sigma_x  \sigma_y  \sigma_c + \ell_1  \sigma_x \sigma_y^2 + 
\ell_2 \sigma_x \sigma_c^2 + \frac{1}{2}\left(2\ell_1+\ell_2\right)  \sigma_x^2, \\
h_y &=& m^2 \sigma_y - \frac{\mathscr{C}}{2}  \sigma_x^2  \sigma_c + \ell_1  \sigma_x^2  \sigma_y + 
\ell_2  \sigma_y  \sigma_c^2 + \left(\ell_1+\ell_2\right)  \sigma_y^2, \\
h_c &=& m^2 \sigma_c - \frac{\mathscr{C}}{2}  \sigma_x^2  \sigma_y  + \ell_1  \sigma_x^2  \sigma_c + 
\ell_2  \sigma_y^2  \sigma_c + \left(\ell_1+\ell_2\right)  \sigma_{c}^{3},
\eea

When considering an external field $\Delta$, represented by the Lagrangian term $\Tr[\Delta(L^{\mu\nu}+L^{\mu\nu})]$, we arrive at the result that
\bea
\Delta &=& \sum_{a=0}^{N_f^2-1} h_a \delta_a = h_0 \delta_0 + h_8 \delta_{15} + h_a \delta_{15}, \label{eq:Delta-rslts}\\
 &=& \left(
\begin{tabular}{c c c c}
$\delta_u$ & $0$ & $0$ & $0$ \\
$0$ & $\delta_d$ & $0$ & $0$ \\
$0$ & $0$ & $\delta_s$ & $0$ \\
$0$ & $0$ & $0$ & $\delta_c$
\end{tabular}
\right),
\eea
from which we deduce that
\bea
 \left(
\begin{tabular}{c}
$\delta_u$ \\
$\delta_d$  \\
$\delta_s$  \\
$\delta_c$
\end{tabular}
\right) &=& \left(
\begin{tabular}{c}
$m_u^2$ \\
$m_d^2$  \\
$m_s^2$  \\
$m_c^2$
\end{tabular}
\right).
\eea
$N_f$ gives the number of quark flavors.

Through the application of the electromagnetic field, $A_{\mu} = g A_{\mu}^a \lambda^a/2$, we derive vector and axialvector meson nonets, 
\bea
L^{\mu\nu} &=& \delta_{\mu}L^{\nu} - i e A^{\mu}\left[T_3, L^{\nu}\right] - \left\{\delta^{\nu}\mathcal{L}^{\mu}-i e A^{\nu}\left[T_3,L^{\mu}\right]\right\}, \\
R^{\mu\nu} &=& \delta_{\mu}R^{\nu} - i e A^{\mu}\left[T_3, R^{\nu}\right] - \left\{\delta^{\nu}R^{\mu}-i e A^{\nu}\left[T_3,R^{\mu}\right]\right\}, 
\eea
where $L^{\mu}=\sum_{a=0}^{N_f^2-1} T_a (V_a^{\mu} + A_a^{\mu})$ and $R^{\mu}=\sum_{a=0}^{N_f^2-1} T_a (V_a^{\mu} - A_a^{\mu})$. The nonvanishing external field matrices $H$ and $\epsilon$ clearly break the chiral symmetry
\bea
H &=& \sum_{a=0}^{N_f^2-1} h_a T_a = h_0 T_0 + h_8 T_8 + h_{15} T_{15}, \label{eq:H-rslts}\\
\epsilon &=& \epsilon_c= m_c^2=\frac{1}{2} \left[m_{\chi_{c0}}^2-m_0^2 - \ell_1\left(\sigma_x^2+\sigma_y^2\right) - 3 \sigma_c^2 \left(\ell_1+\ell_2\right)\right]. \label{eq:epsilon-rslts}
\eea
The generators of the group U($N_f$) are $T_a=\lambda_a/2$, where $\lambda_a$ are the Gell-Mann matrices.

In the isospin-symmetric approximation, it is possible to assign the values $\delta_u=\delta_d=0$. Consequently, for the parameters $\delta_x$, $\delta_y$, and $\delta_c$, one may utilize the mass equations of vector mesons, such as $m^2_{\omega_N}$, $m^2_{\omega_S}$, and $m^2_{\chi_{c1}}$. Then, we get
\bea
\delta_x &=& \frac{1}{2} \left[m^2_{\omega_N}-m_1^2+m_0^2-\frac{\sigma_x^2}{2}\left(h_1+h_2+h_3\right) -\frac{h_1}{2} \left(\sigma_y^2+\sigma_c^2\right) \right], \\ 
\delta_y &=& \frac{1}{2} \left[m^2_{\omega_S}-m_1^2+m_0^2-\frac{\sigma_y^2}{2}\left(\frac{h_1}{2}+h_2+h_3\right) - \frac{h_1}{2} \left(\sigma_x^2+\sigma_c^2\right) \right], \\ 
\delta_c &=& \frac{1}{2} \left[m^2_{\chi_{c1}}-m_1^2+m_0^2 - 2 g_1^2 \sigma_c^2 - \sigma_c^2\left(\frac{h_1}{2}+h_2-h_3\right) -\frac{h_1}{2} \left(\sigma_y^2+\sigma_y^2\right) \right].  
\eea

The other mass parameters are given as follows:
\bea
m^2 &=& m_{\pi}^2 - \frac{f_{\pi}^2}{2} \ell_2 + \mathscr{C}\left[f_K-\frac{f_{\phi}}{2}\right]- \ell_1\left[f_K-\frac{f_{\pi}}{2}\right]^2, \\
m^2_0 &=& \frac{1}{2} \left[m_{a_0}^2+ m_{\sigma_s}^2 - \ell_2\left(\frac{3}{2} \sigma_x^2 + 3 \sigma_y^2\right)\right].
\eea

It is important to note that the parameters $h_1$, $h_2$, and $h_3$ are connected to the quark condensates $\sigma_u$, $\sigma_d$, and $\sigma_s$, which can be formulated in relation to $\sigma_0$, $\sigma_3$, and $\sigma_8$,
\bea
\sigma_u &=& \sqrt{2} \sigma_0 + \sigma_3 + \sigma_8, \\
\sigma_d &=& \sqrt{2} \sigma_0 - \sigma_3 + \sigma_8, \\
\sigma_s &=& \sigma_0 - \sqrt{2} \sigma_8.
\eea
Accordingly, $h_0$, $h_3$, and $h_8$ are specified as
\bea
h_0  &=& \frac{1}{\sqrt{6}} \left[f_{\pi} m_{\pi}^2 + 2 f_K m_k^2\right], \\
h_3  &=& \left[m^2 + \frac{\mathscr{C}}{\sqrt{6}} \sigma_0 -  \frac{\mathscr{C}}{\sqrt{6}} \sigma_8 + \ell_1 \left(\sigma_0^2+\sigma_3^2+\sigma_8^2\right) \right. \nn \\
&& \left. + \ell_2 \left(\sigma_0^2 + \frac{\sigma_3^2}{2} +  \frac{\sigma_8^2}{2} + \sqrt{2} \sigma_0 \sigma_8\right)
\right] \sigma_3, \\
h_8  &=& \frac{2}{3} \left[f_{\pi} m_{\pi}^2 - 2 f_K m_k^2\right].  
\eea

In SU(4)$_{\ell}\, \times $SU(4)$_r$ model, the quark condensates are given as
\bea
\sigma_x &=& \frac{\sigma_0}{\sqrt{2}} + \frac{\sigma_8}{\sqrt{3}} + \frac{\sigma_{15}}{\sqrt{6}}, \\
\sigma_y &=& \frac{\sigma_0}{2} - \sqrt{\frac{2}{3}} \sigma_8 + \frac{1}{2\sqrt{3}} \sigma_{15}, \\
\sigma_c &=& \frac{\sigma_0}{2}  - \frac{\sqrt{3}}{2} \sigma_{15},
\eea
where $\sigma_x$ refers the the condensate of light quarks, i.e., nondegeneratee up and down quarks, whereas $\sigma_y$ ($\sigma_c$) represents the strange (charm) quark condensate. The complex matrix of dimensions $N_f \times N_f$ is associated with scalar $J^{PC}=0^{++}$, pseudoscalar $J^{PC}=0^{-+}$, vector $J^{PC}=0^{--}$, and axialvector mesons $J^{PC}=0^{++}$ \cite{Ahmadov:2023mmy}. By applying the formula $m_0=(g/2)\Phi$, where $g$ represents the Yukawa coupling, it is possible to relate the quark masses to the quark condensates.
\bea
m_u &=& \frac{g}{2}\left[\frac{\sigma_0}{\sqrt{2}} + \frac{\sigma_8}{\sqrt{3}}  + \frac{\sigma_{15}}{\sqrt{6}}\right]= \frac{g}{2} \sigma_x, \\
m_d &=& \frac{g}{2}\left[\frac{\sigma_0}{\sqrt{2}} + \frac{\sigma_8}{\sqrt{3}}  + \frac{\sigma_{15}}{\sqrt{6}}\right]= \frac{g}{2} \sigma_x, \\
m_s &=& \frac{g}{2}\left[\frac{\sigma_0}{\sqrt{2}} - \frac{2\sigma_8}{\sqrt{3}}  + \frac{\sigma_{15}}{\sqrt{6}}\right]=\frac{g}{\sqrt{2}} \sigma_y, \\
m_c &=& \frac{g}{2}\left[\frac{\sigma_0}{\sqrt{2}} - \sqrt{\frac{3}{2}} \sigma_{15}\right]=\frac{g}{\sqrt{2}} \sigma_c.
\eea

At finite temperature, $\sigma_x\rightarrow \sigma_x(T)$, $\sigma_y\rightarrow \sigma_y(T)$, $\sigma_c\rightarrow \sigma_c(T)$. Additionally, it is required that all eLSM parameters be formulated as functions of temperature. As elaborated in section \ref{sec:su3}, the masses of different meson states can be obtained from the second derivative of the grand potential with respect to their respective mass fields, \eqref{eq:GrndPtnSU3}. For SU(4), $a, b \in [0,1,\cdots,15]$. Below is a compilation of the masses associated with noncharmed meson states.

\begin{itemize}%
\item Pseudoscalar charmed mesons
\bea
m_{\pi}^2(T) &=& Z_{\pi}^2(T)\left[m_0^2\left(1 + \frac{T^2}{T^2_0}\right) + \left(\ell_1(T) + \frac{\ell_2(T)}{2} \right)\sigma_x^2(T)+\ell_1(T) \sigma_y^2(T) \right. \nn \\
&&\left. + \ell_1(T) \sigma_c^2(T)\right], \\
m_{K}^2(T) &=& Z_{K}^2(T)\left[m_0^2\left(1 + \frac{T^2}{T^2_0}\right) + \left(\ell_1(T)+\frac{\ell_2(T)}{2}\right)\sigma_x^2(T)-\frac{\ell_2(T)}{\sqrt{2}} \sigma_x(T) \sigma_y(T) \right.\nn \\
&&\left. + \ell_1(T) \left[\sigma_y^2(T)+\sigma_c^2(T)\right] + \ell_2(T) \sigma_y^2(T)
\right], \\
m_{{\eta}_{N}}^2(T) &=& Z_{\pi}^2(T)\left[m_0^2\left(1 + \frac{T^2}{T^2_0}\right) + \left(\ell_1(T)+\frac{\ell_2(T)}{2}\right)\sigma_x^2(T)+\ell_1(T) \left[\sigma_y^2(T)+ \sigma_c^2(T)\right] \right.\nn \\
&&\left. + \frac{\mathscr{C}}{2} \sigma_x^2(T)  \sigma_y^2(T) \sigma_c^2(T) \right], \\
%
%
m_{{\eta}_{S}}^2(T) &=& Z_{{\eta}_{S}}^2(T) \left[m_0^2\left(1 + \frac{T^2}{T^2_0}\right) + \ell_1(T)\left(\sigma_x^2(T)+\sigma_c^2(T)\right) + \left[\ell_1(T)+\ell_2(T)\right] \sigma_y^2(T) \right.\nn \\
&&\left. + \frac{C}{8} \sigma_x^2(T)\sigma_c^2(T) \right].
\eea
%
\item Scalar charmed mesons
\bea
m_{a_{0}}^2(T) &=& m_{0}^2\left(1 + \frac{T^2}{T^2_0}\right) + \ell_1(T)\left[\sigma_x^2(T)+\sigma_y^2(T)+\sigma_c^2(T)\right] + \frac{3\ell_2(T)}{2} \sigma_x^2(T), \\
m_{k^*_{0}}^2(T) &=& Z_{k^*_{0}}^2(T)\left[m_0^2\left(1 + \frac{T^2}{T^2_0}\right) + \left[\ell_1(T)+\frac{\ell_2(T)}{2}\right]\sigma_x^2(T) 
+ \frac{\ell_2(T)}{\sqrt{2}}\sigma_x(T)\sigma_y(T) \right.\nn \\
&&\left. + \ell_1(T)\left[\sigma_y^2(T)+\sigma_c^2(T)\right] + \ell_2(T) \sigma_y^2(T)
\right],  \\
m_{{\sigma}_{N}}^2(T) &=& m_{0}^2\left(1 + \frac{T^2}{T^2_0}\right) + 3\left[\ell_1(T)+\frac{\ell_2(T)}{2}\right]\sigma_x^2(T)+\ell_1(T)\left[\sigma_y^2(T)+\sigma_c^2(T)\right], \hspace*{7mm}\\
m_{{\sigma}_{S}}^2(T) &=& m_{0}^2\left(1 + \frac{T^2}{T^2_0}\right) + \ell_1(T)\left[\sigma_x^2(T)+\sigma_c^2(T)\right] + 3 \left[\ell_1(T)+\ell_2(T)\right] \sigma_y^2(T).
\eea

%
\item Vector charmed mesons
\bea
m^2_{\omega_N}(T) &=& m_1^2(T) - m_0^2\frac{T^2}{T^2_0}  + \frac{1}{2}\left[h_1(T)+h_2(T)+h_3(T)\right] \sigma_x^2(T) \nn \\
&& + \frac{h_1(T)}{2}\left[\sigma_y^2(T)+\sigma_c^2(T)\right] + 2 \delta_x(T), \\
m^2_{\omega_S}(T) &=& m_1^2(T) - m_0^2\frac{T^2}{T^2_0}  + \frac{h_1(T)}{2} \left[\sigma_x^2(T) + \sigma_c^2(T)\right] \nn \\
&& + \left[\frac{h_1(T)}{2}+h_2(T)+h_3(T)\right] \sigma_y^2(T) +  2 \delta_x(T),  \\
m^2_{K^*}(T) &=& m_1^2(T) - m_0^2\frac{T^2}{T^2_0} + \frac{\sigma_x^2(T)}{4}  \left[g_1^2(T) + 2 h_1(T) + h_2(T)\right] \nn \\
&& +\frac{\sigma_x(T)}{\sqrt{2}} \sigma_x(T)\sigma_y(T)\left[h_3(T) - g_1^2(T)\right] + \frac{\sigma_y^2(T)}{2}  \left[g_1^2(T) + h_1(T) + h_2(T)\right] \nn \\
&& + h_1(T) \frac{\sigma_c^2(T)}{2}  + \delta_x(T) + \delta_y(T), \\
m^2_{\rho}(T) &=& m^2_{\omega_N}(T).
\eea

\item Axialvector charmed mesons
\bea
m^2_{a_1}(T) &=&  m_1^2(T) - m_0^2\frac{T^2}{T^2_0} + \frac{h_1(T)}{2} \left[\sigma_y^2(T)+\sigma_c^2(T)\right] \nn \\
&& + \frac{\sigma_x^2(T)}{2} \left[h_1(T)+h_2(T)-h_3(T)\right]+2\delta_x(T), \\
m^2_{f_{1s}}(T) &=&  m_1^2(T) - m_0^2\frac{T^2}{T^2_0}  + \frac{h_1(T)}{2}  \left[\sigma_x^2(T) + \sigma_c^2(T)\right]  \nn \\
&& + \left[\frac{h_1(T)}{2} + h_2(T) - h_3(T)\right] \sigma_y^2(T) + 2 \delta_y(T), \\
m^2_{k_{1}}(T) &=&  m_1^2(T) - m_0^2\frac{T^2}{T^2_0}  + \frac{1}{4}\sigma_x^2(T)\left[g_1^2(T) + 2h_1(T)+h_2(T)\right] \nn \\
&+& \frac{\sigma_y^2(T)}{2}\left[g_1^2(T) + h_1(T) + h_2(T)\right] - \frac{1}{\sqrt{2}} \sigma_x(T) \sigma_y(T) \left[g_1^2(T)-h_3(T)\right] \nn \\
&+& \frac{h_1(T)}{2} \sigma_c^2(T) + \delta_x(T) + \delta_y(T), \\
m^2_{1N}(T) &=& m^2_{a_1}(T).
\eea
\end{itemize}

These sixteen meson states can be systematically analyzed in relation to the noncharmed meson states outlined in section \ref{sec:su3}. We will now introduce the analytical expressions that describe the masses associated with charmed meson states.

\begin{itemize}%
\item Pseudoscalar charmed mesons
\bea
m_D^2(T) &=& Z_D^2(T)\left[m_0^2\left(1 + \frac{T^2}{T^2_0}\right)+\left(\ell_1(T)+\frac{\ell_2(T)}{2}\right)\sigma_x^2(T) + \ell_1(T)\sigma_y^2(T) \right. \nn \\
&& \left. +\left[\ell_1(T)+\ell_2(T)\right]\sigma_c^2(T) - \frac{\ell_2(T)}{\sqrt{2}} \sigma_x(T) \sigma_c(T) + \epsilon_c(T)
\right], \\
m_{\eta_c}^2(T) &=& Z_{\eta_c}^2(T)\left[m_0^2\left(1 + \frac{T^2}{T^2_0}\right)+\lambda_1\left[\sigma_x^2(T) + \sigma_y^2(T)\right]+\left(\ell_1(T)+\ell(T)_2\right)\sigma_c^2(T) \right. \nn \\
&&\left. + \frac{\mathscr{C}}{8}\sigma_x^2(T) \sigma_y^2(T) + 2\epsilon_c(T) \right], \\
%
m_{D_s}^2(T) &=& Z_{D_s}^2\left[m_0^2\left(1 + \frac{T^2}{T^2_0}\right)+\ell_1(T) \sigma_x^2(T) + \left[\ell_1(T)+\ell_2(T)\right]\sigma_y^2 \right. \nn \\
&&\left. +\left[\ell_1(T)+\ell_2(T)\right]\sigma_c^2(T) - \ell_2(T)\sigma_y(T)\sigma_c(T) + \epsilon_c(T)
\right].
\eea

%
\item Scalar charmed mesons
\bea
m_{{\chi}_{c0}}^2(T) &=& m_0^2\left(1 + \frac{T^2}{T^2_0}\right)+\ell_1(T) \left[\sigma_x^2(T)+\sigma_y^2(T) \right] + 3\left[\ell_1(T) +\ell_2(T)\right]\sigma_c^2(T) \nn \\
&& + 2\epsilon_c(T), \\
m_{{D}_{0}^*}^2(T) &=& Z_{D_0^*}^2\left[m_0^2\left(1 + \frac{T^2}{T^2_0}\right)+\left(\ell_1(T)+\frac{\ell_2(T)}{2}\right) \sigma_x^2(T)+\ell_1(T)\sigma_y^2(T) \right. \nn \\
&&\left. + \frac{\ell_2(T)}{\sqrt{2}}\sigma_x(T)\sigma_c(T) + \left[\ell_1(T)+\ell_2(T)\right]\sigma_c^2(T)+\epsilon_c(T)
\right], \\
m_{{D}_{0}^{*0}}^2(T) &=& Z_{D_0^{*0}}^2\left[m_0^2\left(1 + \frac{T^2}{T^2_0}\right)+\left(\ell_1(T)+\frac{\ell_2(T)}{2}\right) \sigma_x^2(T)+\ell_1(T)\sigma_y^2(T) \right. \nn \\
&&\left. + \frac{\ell_2(T)}{\sqrt{2}}\sigma_x(T)\sigma_c(T) + \left[\ell_1(T)+\ell_2(T)\right]\sigma_c^2(T)+\epsilon_c(T)
\right],\\
m_{{D}_{s0}^{*}}^2(T) &=& Z_{D_{s0}^{*}}^2\left[m_0^2\left(1 + \frac{T^2}{T^2_0}\right)+\ell_1(T)\sigma_x^2(T) + \left[\ell_1(T)+\ell_2(T)\right] \sigma_y^2(T) \right. \nn \\
&&\left. + \ell_2(T)\sigma_y(T)\sigma_c(T)+\left[\ell_1(T)+\ell_2(T)\right]\sigma_c^2(T)+\epsilon_c(T)
\right].
\eea

%
\item Vector charmed mesons
\bea
m_{D^{*}}^2(T) &=& m_1^2(T)-m_0^2\frac{T^2}{T^2_0}+\left(\frac{g_1^2(T)}{2}+h_1(T)+\frac{h_2(T)}{2}\right)\frac{\sigma_x^2(T)}{2} \nn \\
&& + \frac{1}{\sqrt{2}} \sigma_x(T)\sigma_c(T)\left[h_3(T)-g_1^2(T)\right] + \frac{1}{2}\left(g_1^2(T)+h_1(T)+h_2(T)\right)\sigma_c^2(T) \nn \\
&& +h_1(T)\frac{\sigma_y^2(T)}{2} +\delta_x(T)+\delta_c(T),  \\
m_{J/\psi}^2(T) &=& m_1^2(T)-m_0^2\frac{T^2}{T^2_0}+\frac{h_1(T)}{2} \left[\sigma_x^2(T)+\sigma_y^2(T)\right] \nn \\
&& +\left(\frac{h_1(T)}{2}+h_2(T)+h_3(T)\right)\sigma_c^2(T) + 2\delta_c(T), \\
m_{D_s^*}^2(T) &=& m_1^2(T)-m_0^2\frac{T^2}{T^2_0}+\frac{1}{2}\left(g_1^2(T)+h_1(T)+h_2(T)\right) \left[\sigma_y^2(T)+\sigma_c^2(T)\right] \nn \\
&& +\frac{h_1(T)}{2}\sigma_x^2(T) + \left(h_3(T)-g_1^2(T)\right)\sigma_y(T)\sigma_c(T) + \delta_y(T) + \delta_c(T).
\eea

\item Axialvector charmed mesons
\bea
m_{D_{s1}}^2(T) &=& m_1^2(T)-m_0^2\frac{T^2}{T^2_0}+\frac{1}{2}\left(g_1^2(T)+h_1(T)+h_2(T)\right) \left[\sigma_y^2(T)-\sigma_c^2(T)\right] \nn \\
&&-\sigma_y(T)\sigma_c(T)\left(g_1^2(T)+h_3(T)\right) + \frac{h_1(T)}{2}\sigma_x^2(T)+\delta_y(T) + \delta_c(T), \\
m_{D_1}^2(T) &=& m_1^2(T)-m_0^2\frac{T^2}{T^2_0}+\frac{1}{4}\left(g_1^2(T)+2h_1(T)+h_2(T)\right) \sigma_x^2(T) \nn \\
&& +\frac{1}{2}\left(g_1^2(T)+h_1(T)+h_2(T)\right) \sigma_c^2(T) - \frac{1}{\sqrt{2}}\left(g_1^2(T)+h_3(T)\right)\sigma_x(T)\sigma_c(T) \nn \\
&& + h_1(T) \frac{\sigma_y^2(T)}{2} + \delta_x(T) + \delta_c(T), \\
m_{{\chi}_{c1}}^2(T) &=& m_1^2(T)-m_0^2\frac{T^2}{T^2_0}+\frac{h_1(T)}{2} \left[\sigma_x^2(T)+\sigma_y^2(T)\right] \nn \\
&&+ \left[\frac{h_1(T)}{2}+h_2(T)-h_3(T)\right] \sigma_c^2(T)  + 2\delta_c(T).
\eea

\end{itemize}

The wavefunction renormalization factors that are now emerging in the previous expressions are given as follows:
\bea
Z_{k_s}(T) &=& \frac{2 m_{K}(T)}{\sqrt{4 m_{K}^2(T)-g_1^2(T) \left[\sigma_x(T)+\sqrt{2}\sigma_y(T)\right]^2}},\\
Z_{\eta_c}(T) &=& \frac{m_{\chi_{c1}}(T)}{\sqrt{m_{\chi_{c1}}^2(T) - 2 g_1^2(T) \sigma_c^2(T)}}, \\ 
Z_D(T) &=& \frac{2 m_{D_1}(T)}{\sqrt{4 m_{D_1}(T) - g_1^2(T) \left[\sigma_c(T) + \sqrt{2}\sigma_c(T)\right]^2}}, \\
Z_{D_s}(T) &=& \frac{\sqrt{2} m_{D_{s_1}}(T)}{\sqrt{2 m_{D_{s_1}}^2(T) - g_1^2(T) \left[\sigma_y^2(T) + \sigma_c(T)\right]^2}}, \\ 
Z_{D_0^*}(T) &=& \frac{2 m_{D^*}(T)}{\sqrt{4 m_{D^*}(T) - g_1^2(T) \left[\sigma_x^2(T) - \sqrt{2}\sigma_c(T)\right]^2}}, \\
Z_{D_0^{*0}}(T) &=& \frac{2 m_{D^{*0}}(T)}{\sqrt{4 m_{D^{*0}}^2(T) - g_1^2(T) \left[\sigma_x(T) - \sqrt{2}\sigma_c(T)\right]^2}}, \\
Z_{D_{s0}^{*}}(T) &=& \frac{ \sqrt{2} m_{D^{*}_{s}}(T) }{
\sqrt{2 m_{D^{*}_{s}}^2(T) - g_1^2(T) \left[\sigma_y(T) - \sigma_c(T)\right]^2}
}.
\eea


The section that follows is devoted to the final conclusions and outlook.

\section{Conclusions and Outlook} 
\label{sec:cncl}

Through the incorporation of mesonic contributions within the eLSM potential, considering both three and four quark flavors, we have introduced an analytical methodology for the evaluation of the masses associated with diverse meson states. We have derived meson states expressed as $\langle \bar{q} q\rangle = \langle \bar{q}_{\ell} q_{r} - \bar{q}_{\ell} q_{r}\rangle \neq 0$ from the effective Lagrangian of the extended linear-sigma model, considering both scenarios with and without charm quarks. In terms of their quantum numbers, including orbital angular momentum $J$, parity $P$, and charge conjugation $C$, these meson states can be classified into several categories: pseudoscalar states with quantum numbers $J^{PC}=0^{-+}$, scalar states with $J^{PC}=0^{++}$, vector states with $J^{PC}=1^{--}$, and axialvector states with $J^{PC}=1^{++}$. We have introduced analytical expressions for the mass spectrum of seventeen non-charmed and twenty-nine charmed meson states at finite temperature. 

The present analytical analyses serve to establish a groundwork for the modification of meson masses in a thermal medium. This investigation is dedicated to the thermal environment and considers the dependence of forty-six mesons, both with and without charm quarks, on finite temperature. As a future direction, we aim to assess the temperature dependence of meson masses through numerical methods. Furthermore, the impact of finite chemical potential on these meson states could be the subject of subsequent research.

\acknowledgments{The authors would like to recognize the support received from the Egyptian Academy for Scientific Research and Technology (ASRT), provided through the ASRT/JINR joint projects (Call 2023), which has made it possible to conduct this research at the Joint Institute for Nuclear Research (JINR) in Russia Federation.}

\conflictsofinterest{The authors declare no conflicts of interest.}


\funding{This research received no external funding.}

\dataavailability{The authors confirm that the data supporting the findings of this study are available within the article.} 

\appendixstart
\appendix
\section{Misprints in  \cite{Ahmadov:2023mmy}}
The publication \cite{Ahmadov:2023mmy} was released in the MDPI Particles journal. It includes a number of printing mistakes. The following is a complete list of all such misprints.
\bea
\mathtt{Eq.}(16) &\rightarrow & \frac{\sigma_0}{2} - \frac{\sqrt{3}}{s} \sigma_{15}, \nn\\
\mathtt{Eq.}(30) &\rightarrow & Z^2_K\left[m_0^2+\left(\lambda_1+\frac{\lambda_2}{2}\right)
\sigma^2_x - \frac{\lambda_2}{\sqrt{2}} \sigma_x \sigma_y + \lambda_1\left(\sigma^2_y+\sigma^2_c\right) + \lambda_2\sigma_y^2\right], \nn \\
\mathtt{Eq.}(34) &\rightarrow & Z^2_{K_0^{\ast}}\left[m_0^2+\left(\lambda_1+\frac{\lambda_2}{2}\right)
\sigma^2_x + \frac{\lambda_2}{\sqrt{2}} \sigma_x \sigma_y + \lambda_1\left(\sigma^2_y+\sigma^2_c\right) + \lambda_2\sigma_y^2\right], \nn \\
\mathtt{Eq.}(41) &\rightarrow &  m_1^2 - m_0^2  + \xcancel{g_1^2 \sigma_x^2} + \frac{1}{2} h_1 \left[\sigma_y^2+\sigma_c^2\right] +  \frac{1}{2} \sigma_x^2\left[h_1+h_2+h_3\right]+2\delta_x,\nn \\
\mathtt{Eq.}(42) &\rightarrow & m_1^2 - m_0^2  + \frac{1}{2} h_1 \left[\sigma_x^2 + \sigma_c^2\right] + \xcancel{2 g_1^2 \sigma_c^2} + \frac{1}{2}\left[h_1 + 2h_2 -2h_3\right] \sigma_y^2 + 2 \delta_y, \nn \\
\mathtt{Eq.}(43) &\rightarrow & m_1^2 - m_0^2 + \frac{1}{4} \sigma_x^2\left(g_1^2+2h_1+h_2\right) + \frac{1}{2} \sigma_y^2\left(g_1^2+h_1+h_2\right) - \frac{1}{\sqrt{2}} \sigma_x\sigma_y\left(g_1^2+h_3\right) \nn \\
&& + \frac{1}{2} h_1 \sigma_c^2 + \delta_x + \sigma_y, \nn \\
\mathtt{Eq.}(45) &\rightarrow & Z^2_{D}\left[m_0^2+\left(\lambda_1+\frac{\lambda_2}{2}\right)
\sigma^2_x + \lambda_1\sigma^2_y + \left(\lambda_1+\lambda_2\right) \sigma_c^2 - \frac{\lambda_2}{2}\sigma_x \sigma_c + \epsilon_c\right], \nn \\
\mathtt{Eq.}(52) &\rightarrow & m_1^2 - m_0^2 + \frac{1}{4} \left(g_1^2+2h_1+h_2\right)\sigma^2_x + \frac{1}{\sqrt{2}} \sigma_x \sigma_c\left(h_3-g_1^2\right) + \frac{1}{2}\left(g_1^2+h_1+h_2\right)\sigma_c^2 \nn \\
&& + \frac{1}{2}h_1\sigma_y^2 + \delta_x+\delta_c, \nn \\
\mathtt{Eq.}(55) &\rightarrow & m_1^2 - m_0^2 + \frac{1}{2} \left(g_1^2+h_1+h_2\right)\left(\sigma^2_y+\sigma^2_c\right) - \sigma_y \sigma_c\left(h_3+g_1^2\right) + \frac{1}{2}h_1 \sigma^2_x + \delta_y+\delta_c, \nn \\
\mathtt{Eq.}(56) &\rightarrow & m_1^2 - m_0^2 + \frac{1}{4} \left(g_1^2+2h_1+h_2\right) \sigma^2_x + \frac{1}{2}\left(g_1^2+h_1+h_2\right) \sigma^2_c - \frac{1}{\sqrt{2}}\sigma_x \sigma_c\left(h_3+g_1^2\right) \nn \\
&& + \frac{1}{2}h_1 \sigma^2_y + \delta_y+\delta_c, \nn \\
\mathtt{Eq.}(57) &\rightarrow & m_1^2-m_0^2+\frac{1}{2}h_1\left[\sigma_x^2+\sigma_y^2\right] + \xcancel{2 g_1^2 \sigma_c^2} + \frac{1}{2}\left(h_1+2h_2-2h_3\right) \sigma_c^2  + 2\delta_c. \nn 
\eea

\reftitle{References}


\bibliography{Azar-CharmedMesonStates1}

\begin{thebibliography}{999}

\bibitem[Ferreira and Papavassiliou(2023)]{Ferreira:2023fva}
Ferreira, M.N.; Papavassiliou, J.
\newblock {Gauge Sector Dynamics in QCD}.
\newblock {\em Particles} {\bf 2023}, {\em 6},~312--363,
  \href{http://arxiv.org/abs/2301.02314}{{\normalfont
  [arXiv:hep-ph/2301.02314]}}.
\newblock {\url{https://doi.org/10.3390/particles6010017}}.

\bibitem[Gorenstein and Mogilevsky(1988)]{Gorenstein:1988fe}
Gorenstein, M.I.; Mogilevsky, O.A.
\newblock {On a Nonperturbative Pressure Effect in Lattice {QCD}}.
\newblock {\em Z. Phys. C} {\bf 1988}, {\em 38},~161--163.
\newblock {\url{https://doi.org/10.1007/BF01574531}}.

\bibitem[Danzer et~al.(2009)Danzer, Gattringer, Liptak, and
  Marinkovic]{Danzer:2009dk}
Danzer, J.; Gattringer, C.; Liptak, L.; Marinkovic, M.
\newblock {A Study of the sign problem for lattice QCD with chemical
  potential}.
\newblock {\em Phys. Lett. B} {\bf 2009}, {\em 682},~240--245,
  \href{http://arxiv.org/abs/0907.3084}{{\normalfont
  [arXiv:hep-lat/0907.3084]}}.
\newblock {\url{https://doi.org/10.1016/j.physletb.2009.11.004}}.

\bibitem[Radzhabov et~al.(2011)Radzhabov, Blaschke, Buballa, and
  Volkov]{Radzhabov:2010dd}
Radzhabov, A.E.; Blaschke, D.; Buballa, M.; Volkov, M.K.
\newblock {Nonlocal PNJL model beyond mean field and the QCD phase transition}.
\newblock {\em Phys. Rev. D} {\bf 2011}, {\em 83},~116004,
  \href{http://arxiv.org/abs/1012.0664}{{\normalfont
  [arXiv:hep-ph/1012.0664]}}.
\newblock {\url{https://doi.org/10.1103/PhysRevD.83.116004}}.

\bibitem[Contrera et~al.(2014)Contrera, Grunfeld, and
  Blaschke]{Contrera:2012wj}
Contrera, G.A.; Grunfeld, A.G.; Blaschke, D.B.
\newblock {Phase diagrams in nonlocal Polyakov-Nambu-Jona-Lasinio models
  constrained by lattice QCD results}.
\newblock {\em Phys. Part. Nucl. Lett.} {\bf 2014}, {\em 11},~342--351,
  \href{http://arxiv.org/abs/1207.4890}{{\normalfont
  [arXiv:hep-ph/1207.4890]}}.
\newblock {\url{https://doi.org/10.1134/S1547477114040128}}.

\bibitem[Karsch et~al.(2003{\natexlab{a}})Karsch, Redlich, and
  Tawfik]{Karsch:2003vd}
Karsch, F.; Redlich, K.; Tawfik, A.
\newblock {Hadron resonance mass spectrum and lattice QCD thermodynamics}.
\newblock {\em Eur. Phys. J. C} {\bf 2003}, {\em 29},~549--556,
  \href{http://arxiv.org/abs/hep-ph/0303108}{{\normalfont [hep-ph/0303108]}}.
\newblock {\url{https://doi.org/10.1140/epjc/s2003-01228-y}}.

\bibitem[Karsch et~al.(2003{\natexlab{b}})Karsch, Redlich, and
  Tawfik]{Karsch:2003zq}
Karsch, F.; Redlich, K.; Tawfik, A.
\newblock {Thermodynamics at nonzero baryon number density: A Comparison of
  lattice and hadron resonance gas model calculations}.
\newblock {\em Phys. Lett. B} {\bf 2003}, {\em 571},~67--74,
  \href{http://arxiv.org/abs/hep-ph/0306208}{{\normalfont [hep-ph/0306208]}}.
\newblock {\url{https://doi.org/10.1016/j.physletb.2003.08.001}}.

\bibitem[Tawfik(2005{\natexlab{a}})]{Tawfik:2004vv}
Tawfik, A.
\newblock {The Influence of strange quarks on QCD phase diagram and chemical
  freeze-out: Results from the hadron resonance gas model}.
\newblock {\em J. Phys. G} {\bf 2005}, {\em 31},~S1105--S1110,
  \href{http://arxiv.org/abs/hep-ph/0410329}{{\normalfont [hep-ph/0410329]}}.
\newblock {\url{https://doi.org/10.1088/0954-3899/31/6/068}}.

\bibitem[Tawfik(2005{\natexlab{b}})]{Tawfik:2004sw}
Tawfik, A.
\newblock {QCD phase diagram: A Comparison of lattice and hadron resonance gas
  model calculations}.
\newblock {\em Phys. Rev. D} {\bf 2005}, {\em 71},~054502,
  \href{http://arxiv.org/abs/hep-ph/0412336}{{\normalfont [hep-ph/0412336]}}.
\newblock {\url{https://doi.org/10.1103/PhysRevD.71.054502}}.

\bibitem[Coppola et~al.(2018)Coppola, G\'omez~Dumm, and
  Scoccola]{Coppola:2018vkw}
Coppola, M.; G\'omez~Dumm, D.; Scoccola, N.N.
\newblock {Charged pion masses under strong magnetic fields in the NJL model}.
\newblock {\em Phys. Lett. B} {\bf 2018}, {\em 782},~155--161,
  \href{http://arxiv.org/abs/1802.08041}{{\normalfont
  [arXiv:hep-ph/1802.08041]}}.
\newblock {\url{https://doi.org/10.1016/j.physletb.2018.04.043}}.

\bibitem[Carlomagno et~al.(2022{\natexlab{a}})Carlomagno, Gomez~Dumm, Noguera,
  and Scoccola]{Carlomagno:2022inu}
Carlomagno, J.P.; Gomez~Dumm, D.; Noguera, S.; Scoccola, N.N.
\newblock {Neutral pseudoscalar and vector meson masses under strong magnetic
  fields in an extended NJL model: Mixing effects}.
\newblock {\em Phys. Rev. D} {\bf 2022}, {\em 106},~074002,
  \href{http://arxiv.org/abs/2205.15928}{{\normalfont
  [arXiv:hep-ph/2205.15928]}}.
\newblock {\url{https://doi.org/10.1103/PhysRevD.106.074002}}.

\bibitem[Carlomagno et~al.(2022{\natexlab{b}})Carlomagno, Gomez~Dumm,
  Villafa\~ne, Noguera, and Scoccola]{Carlomagno:2022arc}
Carlomagno, J.P.; Gomez~Dumm, D.; Villafa\~ne, M.F.I.; Noguera, S.; Scoccola,
  N.N.
\newblock {Charged pseudoscalar and vector meson masses in strong magnetic
  fields in an extended NJL model}.
\newblock {\em Phys. Rev. D} {\bf 2022}, {\em 106},~094035,
  \href{http://arxiv.org/abs/2209.10679}{{\normalfont
  [arXiv:hep-ph/2209.10679]}}.
\newblock {\url{https://doi.org/10.1103/PhysRevD.106.094035}}.

\bibitem[Nambu and Jona-Lasinio(1961)]{PhysRev.122.345}
Nambu, Y.; Jona-Lasinio, G.
\newblock Dynamical Model of Elementary Particles Based on an Analogy with
  Superconductivity. I.
\newblock {\em Phys. Rev.} {\bf 1961}, {\em 122},~345--358.
\newblock {\url{https://doi.org/10.1103/PhysRev.122.345}}.

\bibitem[Gausterer and Sanielevici(1988)]{Gausterer:1988fv}
Gausterer, H.; Sanielevici, S.
\newblock {Can the Chiral Transition in {QCD} Be Described by a Linear $\sigma$
  Model in Three-dimensions?}
\newblock {\em Phys. Lett. B} {\bf 1988}, {\em 209},~533--537.
\newblock {\url{https://doi.org/10.1016/0370-2693(88)91188-4}}.

\bibitem[Friesen et~al.(2019)Friesen, Kalinovsky, and Toneev]{Friesen:2018ojv}
Friesen, A.V.; Kalinovsky, Y.L.; Toneev, V.D.
\newblock {Strange matter and kaon to pion ratio in the SU(3)
  Polyakov\textendash{}Nambu\textendash{}Jona-Lasinio model}.
\newblock {\em Phys. Rev. C} {\bf 2019}, {\em 99},~045201,
  \href{http://arxiv.org/abs/1808.04179}{{\normalfont
  [arXiv:hep-ph/1808.04179]}}.
\newblock {\url{https://doi.org/10.1103/PhysRevC.99.045201}}.

\bibitem[Kalinovsky and Friesen(2015)]{Kalinovsky:2015kzf}
Kalinovsky, Y.L.; Friesen, A.V.
\newblock {Properties of mesons and critical points in the
  Nambu\textendash{}Jona-Lasinio model with different regularizations}.
\newblock {\em Phys. Part. Nucl. Lett.} {\bf 2015}, {\em 12},~737--743.
\newblock {\url{https://doi.org/10.1134/S1547477115060060}}.

\bibitem[Friesen et~al.(2012)Friesen, Kalinovsky, and Toneev]{Friesen:2011wt}
Friesen, A.V.; Kalinovsky, Y.L.; Toneev, V.D.
\newblock {Effects of Model Parameters in Thermodynamics of the PNJL Model}.
\newblock {\em Int. J. Mod. Phys. A} {\bf 2012}, {\em 27},~1250013,
  \href{http://arxiv.org/abs/1111.3042}{{\normalfont
  [arXiv:hep-ph/1111.3042]}}.
\newblock {\url{https://doi.org/10.1142/S0217751X12500133}}.

\bibitem[Giacosa et~al.(2012)Giacosa, Parganlija, Kovacs, and
  Wolf]{Giacosa2012}
Giacosa, F.; Parganlija, D.; Kovacs, P.; Wolf, G.
\newblock {\em Eur. Phys. J. Web Conf.} {\bf 2012}, {\em 37},~08006,
  \href{http://arxiv.org/abs/1208.6202 [hepph]}{{\normalfont [1208.6202
  [hepph]]}}.

\bibitem[Tawfik et~al.(2019)Tawfik, Diab, Ghoneim, and Anwer]{Tawfik:2019tkp}
Tawfik, A.N.; Diab, A.M.; Ghoneim, M.T.; Anwer, H.
\newblock {SU(3) Polyakov Linear-Sigma Model With Finite Isospin Asymmetry: QCD
  Phase Diagram}.
\newblock {\em Int. J. Mod. Phys. A} {\bf 2019}, {\em 34},~1950199,
  \href{http://arxiv.org/abs/1904.09890}{{\normalfont
  [arXiv:hep-ph/1904.09890]}}.
\newblock {\url{https://doi.org/10.1142/S0217751X19501999}}.

\bibitem[Tawfik(2023{\natexlab{a}})]{Tawfik:2023egf}
Tawfik, A.N.
\newblock {QCD Phase Structure and In-Medium Modifications of Meson Masses in
  Polyakov Linear-Sigma Model with Finite Isospin Asymmetry}.
\newblock {\em Universe} {\bf 2023}, {\em 9},~276.
\newblock {\url{https://doi.org/10.3390/universe9060276}}.

\bibitem[Tawfik(2023{\natexlab{b}})]{Tawfik:2023all}
Tawfik, A.N.
\newblock {Isospin Symmetry Breaking in Non-Perturbative QCD}.
\newblock {\em Phys. Sci. Forum} {\bf 2023}, {\em 7},~22.
\newblock {\url{https://doi.org/10.3390/ECU2023-14047}}.

\bibitem[Tawfik and Magdy(2015{\natexlab{a}})]{Tawfik:2015uda}
Tawfik, A.N.; Magdy, N.
\newblock {On SU(3) effective models and chiral phase-transition}.
\newblock {\em Adv. High Energy Phys.} {\bf 2015}, {\em 2015},~563428,
  \href{http://arxiv.org/abs/1509.07114}{{\normalfont
  [arXiv:hep-ph/1509.07114]}}.
\newblock {\url{https://doi.org/10.1155/2015/563428}}.

\bibitem[Tawfik and Magdy(2015{\natexlab{b}})]{Tawfik:2015tga}
Tawfik, A.N.; Magdy, N.
\newblock {SU(3) Polyakov linear-\ensuremath{\sigma} model in magnetic fields:
  Thermodynamics, higher-order moments, chiral phase structure, and meson
  masses}.
\newblock {\em Phys. Rev. C} {\bf 2015}, {\em 91},~015206,
  \href{http://arxiv.org/abs/1501.01124}{{\normalfont
  [arXiv:hep-ph/1501.01124]}}.
\newblock {\url{https://doi.org/10.1103/PhysRevC.91.015206}}.

\bibitem[Eshraim et~al.(2015)Eshraim, Giacosa, and Rischke]{Eshraim2015}
Eshraim, W.I.; Giacosa, F.; Rischke, D.H.
\newblock Phenomenology of charmed mesons in the extended Linear Sigma Model.
\newblock {\em Eur. Phys. J. A} {\bf 2015}, {\em 51},~112,
  \href{http://arxiv.org/abs/1405.5861 [hep-ph]}{{\normalfont [1405.5861
  [hep-ph]]}}.

\bibitem[Eshraim et~al.(2020)Eshraim, Fischer, Giacosa, and
  Parganlija]{Eshraim2020}
Eshraim, W.I.; Fischer, C.S.; Giacosa, F.; Parganlija, D.
\newblock Hybrid phenomenology in a chiral approach.
\newblock {\em Eur. Phys. J. Plus} {\bf 2020}, {\em 135},~945,
  \href{http://arxiv.org/abs/2001.06106 [hep-ph]}{{\normalfont [2001.06106
  [hep-ph]]}}.

\bibitem[Gell-Mann and Levy(1960)]{Gell-Mann:1960mvl}
Gell-Mann, M.; Levy, M.
\newblock {The axial vector current in beta decay}.
\newblock {\em Nuovo Cim.} {\bf 1960}, {\em 16},~705.
\newblock {\url{https://doi.org/10.1007/BF02859738}}.

\bibitem[Burkert et~al.(2023)]{Burkert:2022hjz}
Burkert, V.D.;  et~al.
\newblock {Precision studies of QCD in the low energy domain of the EIC}.
\newblock {\em Prog. Part. Nucl. Phys.} {\bf 2023}, {\em 131},~104032,
  \href{http://arxiv.org/abs/2211.15746}{{\normalfont
  [arXiv:nucl-ex/2211.15746]}}.
\newblock {\url{https://doi.org/10.1016/j.ppnp.2023.104032}}.

\bibitem[Lenaghan and Rischke(2000)]{Lenaghan:1999si}
Lenaghan, J.T.; Rischke, D.H.
\newblock {The O(N) model at finite temperature: Renormalization of the gap
  equations in Hartree and large N approximation}.
\newblock {\em J. Phys. G} {\bf 2000}, {\em 26},~431--450,
  \href{http://arxiv.org/abs/nucl-th/9901049}{{\normalfont [nucl-th/9901049]}}.
\newblock {\url{https://doi.org/10.1088/0954-3899/26/4/309}}.

\bibitem[Petropoulos(1999)]{Petropoulos:1998gt}
Petropoulos, N.
\newblock {Linear sigma model and chiral symmetry at finite temperature}.
\newblock {\em J. Phys. G} {\bf 1999}, {\em 25},~2225--2241,
  \href{http://arxiv.org/abs/hep-ph/9807331}{{\normalfont [hep-ph/9807331]}}.
\newblock {\url{https://doi.org/10.1088/0954-3899/25/11/305}}.

\bibitem[Hu(1974)]{PhysRevD.9.1825}
Hu, B.
\newblock Chiral SU(4) and scale invariance.
\newblock {\em Phys. Rev. D} {\bf 1974}, {\em 9},~1825--1834.
\newblock {\url{https://doi.org/10.1103/PhysRevD.9.1825}}.

\bibitem[Schechter and Singer(1975)]{PhysRevD.12.2781}
Schechter, J.; Singer, M.
\newblock SU(4) sigma model.
\newblock {\em Phys. Rev. D} {\bf 1975}, {\em 12},~2781--2790.
\newblock {\url{https://doi.org/10.1103/PhysRevD.12.2781}}.

\bibitem[Geddes(1980)]{PhysRevD.21.278}
Geddes, H.B.
\newblock Spin-zero mass spectrum in the one-loop approximation in a linear
  SU(4) sigma model.
\newblock {\em Phys. Rev. D} {\bf 1980}, {\em 21},~278--289.
\newblock {\url{https://doi.org/10.1103/PhysRevD.21.278}}.

\bibitem[Ding et~al.(2023)Ding, Roberts, and Schmidt]{Ding:2022ows}
Ding, M.; Roberts, C.D.; Schmidt, S.M.
\newblock {Emergence of Hadron Mass and Structure}.
\newblock {\em Particles} {\bf 2023}, {\em 6},~57--120,
  \href{http://arxiv.org/abs/2211.07763}{{\normalfont
  [arXiv:hep-ph/2211.07763]}}.
\newblock {\url{https://doi.org/10.3390/particles6010004}}.

\bibitem[Geddes(1980)]{Geddes:1979nd}
Geddes, H.B.
\newblock {The Spin 0 Mass Spectrum in the One Loop Approximation in a Linear
  SU(4) Sigma Model}.
\newblock {\em Phys. Rev. D} {\bf 1980}, {\em 21},~278.
\newblock {\url{https://doi.org/10.1103/PhysRevD.21.278}}.

\bibitem[Diab et~al.(2016)Diab, Ahmadov, Tawfik, and Dahab]{Diab:2016iig}
Diab, A.M.; Ahmadov, A.I.; Tawfik, A.N.; Dahab, E.A.E.
\newblock {SU($4$) Polyakov linear-sigma model at finite temperature and
  density}.
\newblock {\em PoS} {\bf 2016}, {\em ICHEP2016},~634,
  \href{http://arxiv.org/abs/1611.02693}{{\normalfont
  [arXiv:nucl-th/1611.02693]}}.
\newblock {\url{https://doi.org/10.22323/1.282.0634}}.

\bibitem[Abdel Aal~Diab and Tawfik(2018)]{AbdelAalDiab:2018hrx}
Abdel Aal~Diab, A.M.; Tawfik, A.N.
\newblock {Quark-hadron phase structure of QCD matter from SU($4$) Polyakov
  linear sigma model}.
\newblock {\em EPJ Web Conf.} {\bf 2018}, {\em 177},~09005,
  \href{http://arxiv.org/abs/1801.10234}{{\normalfont
  [arXiv:hep-ph/1801.10234]}}.
\newblock {\url{https://doi.org/10.1051/epjconf/201817709005}}.

\bibitem[Tawfik et~al.(2014)Tawfik, Magdy, and Diab]{Tawfik:2014uka}
Tawfik, A.; Magdy, N.; Diab, A.
\newblock {Polyakov linear SU(3) $\sigma$ model: Features of higher-order
  moments in a dense and thermal hadronic medium}.
\newblock {\em Phys. Rev. C} {\bf 2014}, {\em 89},~055210,
  \href{http://arxiv.org/abs/1405.0577}{{\normalfont
  [arXiv:hep-ph/1405.0577]}}.
\newblock {\url{https://doi.org/10.1103/PhysRevC.89.055210}}.

\bibitem[Tawfik and Diab(2015)]{Tawfik:2014gga}
Tawfik, A.N.; Diab, A.M.
\newblock {Polyakov SU(3) extended linear- \ensuremath{\sigma} model: Sixteen
  mesonic states in chiral phase structure}.
\newblock {\em Phys. Rev. C} {\bf 2015}, {\em 91},~015204,
  \href{http://arxiv.org/abs/1412.2395}{{\normalfont
  [arXiv:hep-ph/1412.2395]}}.
\newblock {\url{https://doi.org/10.1103/PhysRevC.91.015204}}.

\bibitem[Tawfik et~al.(2016)Tawfik, Diab, Ezzelarab, and
  Shalaby]{Tawfik:2016lih}
Tawfik, A.N.; Diab, A.M.; Ezzelarab, N.; Shalaby, A.G.
\newblock {QCD thermodynamics and magnetization in nonzero magnetic field}.
\newblock {\em Adv. High Energy Phys.} {\bf 2016}, {\em 2016},~1381479,
  \href{http://arxiv.org/abs/1604.00043}{{\normalfont
  [arXiv:hep-ph/1604.00043]}}.
\newblock {\url{https://doi.org/10.1155/2016/1381479}}.

\bibitem[Tawfik et~al.(2018)Tawfik, Diab, and Hussein]{Tawfik:2016gye}
Tawfik, A.N.; Diab, A.M.; Hussein, M.T.
\newblock {Quark\textendash{}hadron phase structure, thermodynamics, and
  magnetization of QCD matter}.
\newblock {\em J. Phys. G} {\bf 2018}, {\em 45},~055008,
  \href{http://arxiv.org/abs/1604.08174}{{\normalfont
  [arXiv:hep-lat/1604.08174]}}.
\newblock {\url{https://doi.org/10.1088/1361-6471/aaba9e}}.

\bibitem[Tawfik et~al.(2016{\natexlab{a}})Tawfik, Diab, and
  Hussein]{Tawfik:2016ihn}
Tawfik, A.N.; Diab, A.M.; Hussein, T.M.
\newblock {SU(3) Polyakov linear-sigma model: bulk and shear viscosity of QCD
  matter in finite magnetic field}.
\newblock {\em Int. J. Adv. Res. Phys. Sci.} {\bf 2016}, {\em 3},~4--14,
  \href{http://arxiv.org/abs/1608.01034}{{\normalfont
  [arXiv:hep-ph/1608.01034]}}.

\bibitem[Tawfik et~al.(2016{\natexlab{b}})Tawfik, Diab, and
  Hussein]{Tawfik:2016edq}
Tawfik, A.N.; Diab, A.M.; Hussein, M.T.
\newblock {SU(3) Polyakov linear-sigma model: Conductivity and viscous
  properties of QCD matter in thermal medium}.
\newblock {\em Int. J. Mod. Phys. A} {\bf 2016}, {\em 31},~1650175,
  \href{http://arxiv.org/abs/1610.06041}{{\normalfont
  [arXiv:nucl-th/1610.06041]}}.
\newblock {\url{https://doi.org/10.1142/S0217751X1650175X}}.

\bibitem[Tawfik et~al.(2018)Tawfik, Diab, and Hussein]{Tawfik:2017cdx}
Tawfik, A.N.; Diab, A.M.; Hussein, M.T.
\newblock {SU(3) Polyakov linear-sigma model: Magnetic properties of QCD matter
  in thermal and dense medium}.
\newblock {\em J. Exp. Theor. Phys.} {\bf 2018}, {\em 126},~620--632,
  \href{http://arxiv.org/abs/1712.03264}{{\normalfont
  [arXiv:hep-ph/1712.03264]}}.
\newblock {\url{https://doi.org/10.1134/S1063776118050138}}.

\bibitem[Tawfik et~al.(2019)Tawfik, Diab, and Hussein]{Tawfik:2019rdd}
Tawfik, A.N.; Diab, A.M.; Hussein, M.T.
\newblock {Chiral phase structure of the sixteen meson states in the SU(3)
  Polyakov linear-sigma model for finite temperature and chemical potential in
  a strong magnetic field}.
\newblock {\em Chin. Phys. C} {\bf 2019}, {\em 43},~034103,
  \href{http://arxiv.org/abs/1901.03293}{{\normalfont
  [arXiv:hep-ph/1901.03293]}}.
\newblock {\url{https://doi.org/10.1088/1674-1137/43/3/034103}}.

\bibitem[Tawfik et~al.(2020)Tawfik, Greiner, Diab, Ghoneim, and
  Anwer]{Tawfik:2019kaz}
Tawfik, A.N.; Greiner, C.; Diab, A.M.; Ghoneim, M.T.; Anwer, H.
\newblock {Polyakov linear-$\sigma$ model in mean-field approximation and
  optimized perturbation theory}.
\newblock {\em Phys. Rev. C} {\bf 2020}, {\em 101},~035210,
  \href{http://arxiv.org/abs/1908.05939}{{\normalfont
  [arXiv:hep-ph/1908.05939]}}.
\newblock {\url{https://doi.org/10.1103/PhysRevC.101.035210}}.

\bibitem[Tawfik and Diab(2021)]{Tawfik:2021eeb}
Tawfik, A.N.; Diab, A.M.
\newblock {Chiral magnetic properties of QCD phase-diagram}.
\newblock {\em Eur. Phys. J. A} {\bf 2021}, {\em 57},~200,
  \href{http://arxiv.org/abs/2106.04576}{{\normalfont
  [arXiv:hep-ph/2106.04576]}}.
\newblock {\url{https://doi.org/10.1140/epja/s10050-021-00501-z}}.

\bibitem[Klempt and Zaitsev(2007)]{Klempt:2007cp}
Klempt, E.; Zaitsev, A.
\newblock {Glueballs, Hybrids, Multiquarks. Experimental facts versus QCD
  inspired concepts}.
\newblock {\em Phys. Rept.} {\bf 2007}, {\em 454},~1--202,
  \href{http://arxiv.org/abs/0708.4016}{{\normalfont
  [arXiv:hep-ph/0708.4016]}}.
\newblock {\url{https://doi.org/10.1016/j.physrep.2007.07.006}}.

\bibitem[Amsler and Törnqvist(2004)]{AMSLER200461}
Amsler, C.; Törnqvist, N.A.
\newblock Mesons beyond the naive quark model.
\newblock {\em Physics Reports} {\bf 2004}, {\em 389},~61--117.
\newblock
  {\url{https://doi.org/https://doi.org/10.1016/j.physrep.2003.09.003}}.

\bibitem[Vafa and Witten(1984)]{VAFA1984173}
Vafa, C.; Witten, E.
\newblock Restrictions on symmetry breaking in vector-like gauge theories.
\newblock {\em Nuclear Physics B} {\bf 1984}, {\em 234},~173--188.
\newblock {\url{https://doi.org/https://doi.org/10.1016/0550-3213(84)90230-X}}.

\bibitem[Rosenzweig et~al.(1981)Rosenzweig, Salomone, and
  Schechter]{Rosenzweig:1981cu}
Rosenzweig, C.; Salomone, A.; Schechter, J.
\newblock {A Pseudoscalar Glueball, the Axial Anomaly and the Mixing Problem
  for Pseudoscalar Mesons}.
\newblock {\em Phys. Rev. D} {\bf 1981}, {\em 24},~2545--2548.
\newblock {\url{https://doi.org/10.1103/PhysRevD.24.2545}}.

\bibitem[Ahmadov et~al.(2024)Ahmadov, Alshehri, and Tawfik]{Ahmadov:2023mmy}
Ahmadov, A.I.; Alshehri, A.A.; Tawfik, A.N.
\newblock {Mass Spectrum of Non-Charmed and Charmed Meson States in Extended
  Linear-Sigma Model}.
\newblock {\em Particles} {\bf 2024}, {\em 7},~560--575,
  \href{http://arxiv.org/abs/2310.09387}{{\normalfont
  [arXiv:hep-ph/2310.09387]}}.
\newblock {\url{https://doi.org/10.3390/particles7030031}}.

\bibitem[Bowman and Kapusta(2009)]{Bowman:2008kc}
Bowman, E.S.; Kapusta, J.I.
\newblock {Critical Points in the Linear Sigma Model with Quarks}.
\newblock {\em Phys. Rev. C} {\bf 2009}, {\em 79},~015202,
  \href{http://arxiv.org/abs/0810.0042}{{\normalfont
  [arXiv:nucl-th/0810.0042]}}.
\newblock {\url{https://doi.org/10.1103/PhysRevC.79.015202}}.

\bibitem[Datta and Gupta(2009)]{Datta:2009jn}
Datta, S.; Gupta, S.
\newblock {Scaling and the continuum limit of the finite temperature
  deconfinement transition in SU$(N_c)$ pure gauge theory}.
\newblock {\em Phys. Rev. D} {\bf 2009}, {\em 80},~114504,
  \href{http://arxiv.org/abs/0909.5591}{{\normalfont
  [arXiv:hep-lat/0909.5591]}}.
\newblock {\url{https://doi.org/10.1103/PhysRevD.80.114504}}.

\bibitem[Workman et~al.(2022)]{ParticleDataGroup:2022pth}
Workman, R.L.;  et~al.
\newblock {Review of Particle Physics}.
\newblock {\em PTEP} {\bf 2022}, {\em 2022},~083C01.
\newblock {\url{https://doi.org/10.1093/ptep/ptac097}}.

\bibitem[Gasser and Zarnauskas(2010)]{Gasser:2010wz}
Gasser, J.; Zarnauskas, G.R.S.
\newblock {On the pion decay constant}.
\newblock {\em Phys. Lett. B} {\bf 2010}, {\em 693},~122--128,
  \href{http://arxiv.org/abs/1008.3479}{{\normalfont
  [arXiv:hep-ph/1008.3479]}}.
\newblock {\url{https://doi.org/10.1016/j.physletb.2010.08.021}}.

\bibitem[Sanz-Cillero(2004)]{Sanz-Cillero:2004hed}
Sanz-Cillero, J.J.
\newblock {Pion and kaon decay constants: Lattice versus resonance chiral
  theory}.
\newblock {\em Phys. Rev. D} {\bf 2004}, {\em 70},~094033,
  \href{http://arxiv.org/abs/hep-ph/0408080}{{\normalfont [hep-ph/0408080]}}.
\newblock {\url{https://doi.org/10.1103/PhysRevD.70.094033}}.

\bibitem[Schaefer and Wagner(2009)]{Schaefer:2008hk}
Schaefer, B.J.; Wagner, M.
\newblock {The Three-flavor chiral phase structure in hot and dense QCD
  matter}.
\newblock {\em Phys. Rev. D} {\bf 2009}, {\em 79},~014018,
  \href{http://arxiv.org/abs/0808.1491}{{\normalfont
  [arXiv:hep-ph/0808.1491]}}.
\newblock {\url{https://doi.org/10.1103/PhysRevD.79.014018}}.

\bibitem[Schaefer and Wambach(2007)]{Schaefer:2006ds}
Schaefer, B.J.; Wambach, J.
\newblock {Susceptibilities near the QCD (tri)critical point}.
\newblock {\em Phys. Rev. D} {\bf 2007}, {\em 75},~085015,
  \href{http://arxiv.org/abs/hep-ph/0603256}{{\normalfont [hep-ph/0603256]}}.
\newblock {\url{https://doi.org/10.1103/PhysRevD.75.085015}}.

\bibitem[Scavenius et~al.(2001)Scavenius, Mocsy, Mishustin, and
  Rischke]{Scavenius:2000qd}
Scavenius, O.; Mocsy, A.; Mishustin, I.N.; Rischke, D.H.
\newblock {Chiral phase transition within effective models with constituent
  quarks}.
\newblock {\em Phys. Rev. C} {\bf 2001}, {\em 64},~045202,
  \href{http://arxiv.org/abs/nucl-th/0007030}{{\normalfont [nucl-th/0007030]}}.
\newblock {\url{https://doi.org/10.1103/PhysRevC.64.045202}}.

\bibitem[Cheng et~al.(2008)]{Cheng:2007jq}
Cheng, M.;  et~al.
\newblock {The QCD equation of state with almost physical quark masses}.
\newblock {\em Phys. Rev. D} {\bf 2008}, {\em 77},~014511,
  \href{http://arxiv.org/abs/0710.0354}{{\normalfont
  [arXiv:hep-lat/0710.0354]}}.
\newblock {\url{https://doi.org/10.1103/PhysRevD.77.014511}}.

\bibitem[Kapusta and Gale(2011)]{Kapusta:2006pm}
Kapusta, J.I.; Gale, C.
\newblock {\em {Finite-temperature field theory: Principles and applications}};
  Cambridge Monographs on Mathematical Physics, Cambridge University Press,
  2011.
\newblock {\url{https://doi.org/10.1017/CBO9780511535130}}.

\bibitem[Parganlija et~al.(2013)Parganlija, Kovacs, Wolf, Giacosa, and
  Rischke]{Parganlija:2012fy}
Parganlija, D.; Kovacs, P.; Wolf, G.; Giacosa, F.; Rischke, D.H.
\newblock {Meson vacuum phenomenology in a three-flavor linear sigma model with
  (axial-)vector mesons}.
\newblock {\em Phys. Rev. D} {\bf 2013}, {\em 87},~014011,
  \href{http://arxiv.org/abs/1208.0585}{{\normalfont
  [arXiv:hep-ph/1208.0585]}}.
\newblock {\url{https://doi.org/10.1103/PhysRevD.87.014011}}.

\end{thebibliography}


%



\end{document}